\documentclass[preprint,prd,noshowpacs,nofootinbib]{revtex4}
\usepackage{amssymb}
\usepackage{amsmath}
\usepackage{graphicx}

\begin{document}

\title{Time dependent electromagnetic fields and 4-dimensional Stokes' theorem}

\author{Ryan Andosca}
\email{randosca@mail.fresnostate.edu}
\affiliation{Department of Physics, California State University Fresno, Fresno, CA 93740-8031, USA}

\author{Douglas Singleton}
\email{dougs@csufresno.edu}
\affiliation{Department of Physics, California State University Fresno, Fresno, CA 93740-8031, USA \\
and \\
ICTP South American Institute for Fundamental Research,
UNESP - Univ. Estadual Paulista
Rua Dr. Bento T. Ferraz 271, 01140-070, S{\~a}o Paulo, SP, Brasil}

\date{\today}

\begin{abstract}
Stokes' theorem is central to many aspects of physics -- electromagnetism, the Aharonov-Bohm effect, 
and Wilson loops to name a few. However, the pedagogical examples and research 
work almost exclusively focus on situations where the fields are time-independent so that one need only deal with 
purely spatial line integrals ({\it e.g.} $\oint {\bf A} \cdot d{\bf x}$) and purely spatial area integrals
({\it e.g.} $\int (\nabla \times {\bf A}) \cdot d{\bf a} = \int {\bf B} \cdot d{\bf a}$). Here we address this gap 
by giving some explicit examples of how Stokes' theorem plays out with time-dependent fields in a 
full 4-dimensional spacetime context. We also discuss some unusual features of Stokes' theorem with time-dependent fields
related to gauge transformations and non-simply connected topology.
\end{abstract}

\maketitle
\section{Stokes' theorem in 3 and 4 dimensions}
\subsection{3D Stokes' theorem}
Stokes' theorem is used in many areas of physics, particularly in electricity and  magnetism where
it gives a connection between the electromagnetic potentials ({\it i.e.} $\phi$ and ${\bf A}$)
and the fields ({\it i.e.} ${\bf E}$ and ${\bf B}$). Through Stokes' theorem, the connection between the 
line integral of the 3-vector potential, ${\bf A}$, and the area integral of the magnetic field, ${\bf B}$, is
\begin{equation}
\label{stokes}
\oint _C {\bf A} \cdot d{\bf x} = \int _S (\nabla \times {\bf A}) \cdot d{\bf a} = \int _S {\bf B} \cdot d{\bf a} ~,
\end{equation}
The subscripts $C$ and $S$ on the integrals indicate line and surface integrals respectively. The closed contour $C$
is spanned by an infinite number of possible surfaces $S$. The contour has a direction of traversal which is determined by
the direction of $d {\bf x}$, and this determines the direction of the vector area, $d{\bf a}$, of the surface, $S$, via
the right-hand-rule (wrap the fingers of the right hand in the direction that the contour is traversed and the
thumb points in the direction of the vector area). This issue of the directionality of the area in Stokes' 
theorem will be important (but less familiar) when we move from a purely spatial area to a {\it spacetime}
area. In the rest of the paper we will drop the subscripts $C$ and $S$ if there is no confusion as to whether the integral 
is a line or surface integral. 

The usual pedagogical examples of \eqref{stokes} involve time-independent 3-vector potentials and 
magnetic fields. One common example in cylindrical coordinates is a solenoid of radius $R$, with the 
axis of the solenoid and the magnetic field along the $z$-axis. The magnetic field for this setup is
\begin{equation}
\label{b-field}
{\bf B} = B_0 {\hat {\bf z}} ~~ {\rm for} ~~  \rho <R ~~~~~ {\rm and} ~~~~~  {\bf B} = {\bf 0} ~~ {\rm for} ~~ \rho \ge R ~,
\end{equation}  
where $B_0$ is the constant value of the magnetic field. The vector potential has the following form
\begin{equation}
\label{a-time}
{\bf A} = \frac{B_0 \rho}{2} {\hat {\boldsymbol \varphi}} ~~ {\rm for} ~~ \rho <R ~~~~{\rm and}~~~~~
{\bf A} = \frac{B_0 R^2}{2 \rho} {\hat {\boldsymbol \varphi}} ~~ {\rm for} ~~ \rho \ge R ~.
\end{equation}
For a contour that is a circle of radius $\rho >R$ which goes around the solenoid, the area integral 
in \eqref{stokes} yields $\int {\bf B} \cdot d{\bf a} = B_0 \pi R^2$. For the line integral one 
similarly finds $\oint {\bf A} \cdot d{\bf x} = B_0 \pi R^2$. Thus, for this setup, Stokes' theorem works out to give
\begin{equation}
\label{stokes-ex}
\oint {\bf A} \cdot d{\bf x} = B_0 \pi R^2 = \int {\bf B} \cdot d{\bf a} ~.
\end{equation}

The above example is also the heart of the time-independent Aharonov-Bohm effect \cite{AB} where one performs the usual quantum mechanical
two-slit experiment but with an infinite solenoid, described by the above ${\bf B}$ and ${\bf A}$, inserted between the
slits of the experiment. For example, say one sends electrons at a two-slit set up. The electrons will form an interference
pattern on a screen placed down stream from the two slits due to the quantum mechanical wave nature of the electrons. In this 
simple two-slit experiment the interference occurs due to the phase difference of the wavefunction coming from
the two different slits, and this comes from the path length difference from each slit to whatever point on the 
down stream screen one is interested in. The Aharonov-Bohm effect comes from placing an infinite solenoid between the slits. 
Classically one does not expect any change since, classically, charged particles only respond to the magnetic field, via
$\frac{{\bf v}}{c} \times {\bf B}$, not the vector potential. However quantum mechanically, due to minimal coupling, the 
electrons will pick up a phase, $\frac{e}{\hbar c} \int _\gamma {\bf A} \cdot d{\bf x}$, when traveling along a contour $\gamma$.
Now for some particular point on the screen there will be two paths leading from each slit to that point -- call these two
paths $\gamma_1$ and $\gamma_2$. Now in addition to the phase difference due to the path length difference there will be an
additional phase difference coming from the line integrals of the vector potential namely 
$\frac{e}{\hbar c} \int _{\gamma_1} {\bf A} \cdot d{\bf x}-\frac{e}{\hbar} \int _{\gamma_2} {\bf A} \cdot d{\bf x} =
\frac{e}{\hbar c} \oint {\bf A} \cdot d{\bf x} = \frac{e}{\hbar c}\int {\bf B} \cdot d {\bf a}$. Thus one gets a phase shift of the
standard interference pattern of the two-slit experiment, which is given by
$\frac{e}{\hbar c} \oint {\bf A} \cdot d{\bf x} = \frac{e}{\hbar c}\int {\bf B} \cdot d {\bf a}$. 
This is the heart of the time-independent Aharonov-Bohm experiment --
that one gets a phase shift in the interference pattern of the two-slit experiment despite the fact that the electrons
move in a region which if ${\bf B}$ field free, but where the vector potential, ${\bf A}$, is non-zero. A fuller and more detailed 
account of the Aharonov-Bohm effect can be found in section 3.4 of reference \cite{ryder}. Due to the close connection
between Stokes' theorem and the Aharonov-Bohm effect we have in mind that the contours and surfaces discussed in this
paper in connection with Stokes' theorem should be those associated with the paths and surfaces of particles in an
Aharonov-Bohm experiment.              
 
\begin{figure}
  \centering
	\includegraphics[trim = 0mm 0mm 0mm 0mm, clip, width=15.0cm]{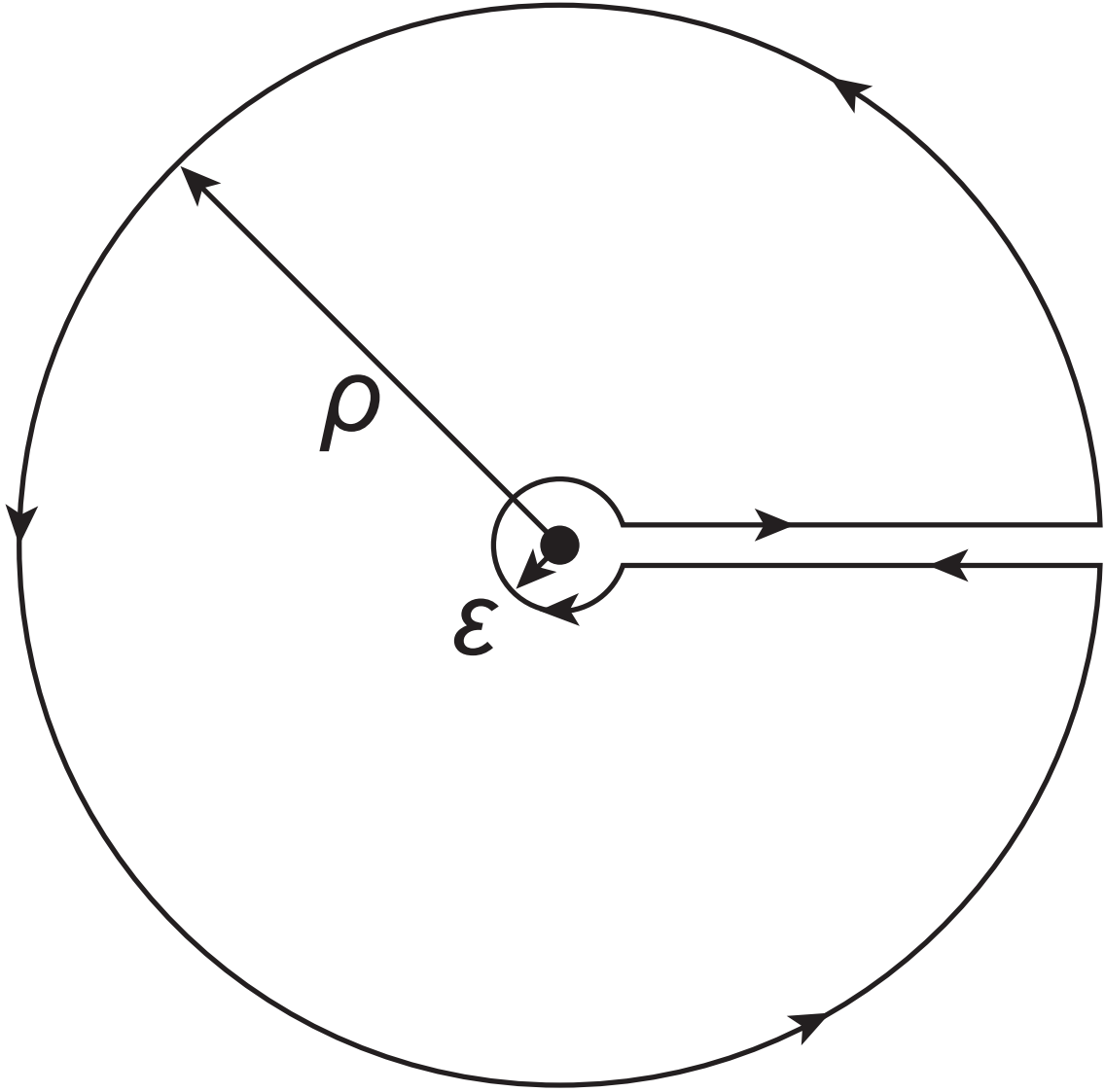}
\caption{{\scriptsize The modified contour for the gauge potential from \eqref{a-time-1} The inner circle has a radius 
$\epsilon \ll R$ and the outer circle a radius $\rho > R $.}}
\label{fig1}
\end{figure}

The magnetic field in \eqref{b-field} can also be obtained from a vector potential of the following form
\begin{equation}
\label{a-time-1}
{\bf A}' = \left[ \frac{B_0 \rho}{2} - \frac{B_0 R^2}{2 \rho} \right] {\hat {\boldsymbol \varphi}} ~~ {\rm for} ~~ \rho <R ~~~~{\rm and}~~~~~
{\bf A}'= {\bf 0} ~~ {\rm for} ~~ \rho \ge R ~.
\end{equation}
This form of the vector potential is related to the original form given in \eqref{a-time} by the following 
gauge transformation
\begin{equation}
\label{g-trans-gen}
{\bf A}'  \rightarrow {\bf A }+ {\bf \nabla} \chi ~~~{\rm with} ~~~ \chi = - \frac{B_0 R^2 \varphi}{2}
\end{equation}
Note that the gauge function, $\chi$ is non-single valued and ${\bf A}'$, for $\rho <R$, has a $\frac{1}{\rho}$ singularity. 
These features (singular vector potential and non-single valued gauge function) indicate that while the ${\bf B}$ field
produced by the two different vector potentials in \eqref{a-time} and \eqref{a-time-1} is the same the physical situation is different --
for the vector potential \eqref{a-time-1} one has the original solenoid of radius $R$ plus an idealized, infinitely thin 
solenoid placed along the symmetry axis with a current flowing in the opposite direction of the original solenoid. 
We will show shortly how to deal with this singularity in ${\bf A}'$. 
One might naively conclude that Stokes' theorem fails in this new gauge. The area integral, $\int {\bf B} \cdot d{\bf a} = B_0 \pi R^2$, 
is still the same since the ${\bf B}$ field is still given by \eqref{b-field}. However, for the circular contour with radius $\rho >R$,
apparently $\oint {\bf A} ' \cdot d{\bf x} = 0$ since ${\bf A}' =0$. The problem is the singularity at $\rho =0$ 
in ${\bf A}'$ inside the solenoid. Due to this ``puncture", the space is said to have a non-simply connected topology. One can not span 
the simple circle contour with a surface that includes $\rho =0$ since this point is no longer part of the space. 
To deal with this ``puncture" at $\rho =0$, we need to deform the simple circle contour into the more 
complex contour in Fig. \eqref{fig1}, which avoids $\rho=0$. The outer circular contour gives zero
($\int_{outer} {\bf A}' \cdot d{\bf x} =0$) since ${\bf A}'=0$ for $\rho >R$. The line integrals for the two
radial segments cancel. The only non-zero contribution is from the inner circular contour with an infinitesimal radius $\epsilon \ll R$.
Since the inner contour is traversed in the opposite direction from the outer contour, the inner line integral 
has a negative sign relative to the outer line integral. Putting it all together one finds that for the contour 
in Fig. \eqref{fig1}.
\begin{eqnarray}
\label{stokes-ex-1}
\oint {\bf A}' \cdot d {\bf x} = \int _{inner} {\bf A} ' \cdot d {\bf x} = 
- \int _0 ^{2 \pi} \left[ \frac{B_0 \epsilon}{2} - \frac{B_0 R^2}{2 \epsilon} \right] \epsilon d \varphi
= B_0 \pi R^2 - B_0 \pi \epsilon ^2 \rightarrow B_0 \pi R^2  ~.
\end{eqnarray}
In the last step we have let $\epsilon \rightarrow 0$. This removes the singularity in ${\bf A}'$
at $\rho = 0$, and we find $\oint {\bf A}' \cdot d {\bf x} = \int {\bf B} \cdot d {\bf a}$ for the contour
in Fig. \eqref{fig1}, thus satisfying Stokes' theorem. 

\subsection{4D Stokes' theorem}

The discussion of the previous subsection was in terms of the 3-vector potential. Since ${\bf A}$ 
is part of the 4-vector $A^\mu = (\phi , {\bf A})$, we can generalize the first expression in equation \eqref{stokes} as
\begin{equation}
\label{stokes2}
\oint {\bf A} \cdot d{\bf x} \rightarrow \oint A_\mu dx ^\mu  = -\oint \phi ~ c dt + \oint {\bf A} \cdot d{\bf x} ~,
\end{equation}   
This generalization was noted already in \cite{AB} and is discussed in 
more detail in references \cite{macdougall, ryder}. Equation \eqref{stokes2} involves both 
a spatial line integral as well as a time integral, thus making it ideal for
time-dependent situations. In the section below we will make this last statement more concrete by looking
at the specific example of a time-dependent solenoid with various contours. It should be
noted that throughout the paper we use the $(-+++)$ metric signature rather than the $(+---)$
signature used in references \cite{AB, macdougall, ryder}. Next, the right hand side of \eqref{stokes}, 
which involves the magnetic field, can be generalized as \cite{macdougall, ryder}
\begin{equation}
\label{stokes3}
\int {\bf B} \cdot d{\bf a} \rightarrow   \int {\bf E} \cdot d{\bf x} ~ c dt + \int {\bf B} \cdot d{\bf a} 
= \frac{1}{2} \int F_{\mu \nu} d \sigma ^{\mu \nu}  ~,
\end{equation}
where in the last expression $F_{\mu \nu} = \partial _\mu A_\nu - \partial _\nu A_\mu$ 
is the Maxwell field strength tensor and $d \sigma ^{\mu \nu}$ is an infinitesimal spacetime area.
The signs of the area integrals of the fields of the intermediate expression in \eqref{stokes3} 
can be checked using the expression for the contour integrals of the potentials in \eqref{stokes2} in the
following way: dropping the scalar potential part of \eqref{stokes2} and the electric area integral of \eqref{stokes3}
one recovers the usual result $\oint {\bf A} \cdot d{\bf x} = \int {\bf B} \cdot d{\bf a}$;
in turn dropping the 3-vector potential line integral of \eqref{stokes2} and magnetic field area integral part of \eqref{stokes3},
one recovers the standard relationship $\phi = - \int {\bf E} \cdot d{\bf x}$ (inside the time integration).

In the next section, we apply Stokes' theorem to an explicit example which involves time-dependent fields and potentials
and which uses the full four-vector potential, $A ^\mu = (\phi, {\bf A} )$.

\section{The infinite, time-dependent solenoid}
\label{solenoid}
In this section we study the case of an infinite solenoid of radius $R$ with a time-dependent magnetic flux. We first consider
a spacetime loop which does not enclose the solenoid. For concreteness and simplicity, we take the time dependence
to be linear so that the 4-vector potential takes the form
\begin{equation}
\label{sole-a}
A^\mu = \left(0, 0, \frac{B_0 t \rho}{2} {\hat {\boldsymbol \varphi}} , 0 \right) ~~ {\rm for} ~~ \rho <R ~~~{\rm and}~~~
A^\mu = \left (0, 0, \frac{B_0 t R^2}{2 \rho} {\hat {\boldsymbol \varphi}} , 0 \right) ~~ {\rm for} ~~ \rho \ge R ~.
\end{equation}
The scalar potential is zero ({\it i.e.} $\phi = 0$).
This is similar to the expression given in \eqref{a-time} but with $B_0 \rightarrow B_0 t$. Note that in \eqref{sole-a} $B_0$ is a
rate of magnetic field strength change, while in \eqref{a-time} $B_0$ is just the magnetic field strength. The magnetic field
connected with the vector potential in \eqref{sole-a} is  
\begin{equation}
\label{sole-b}
{\bf B} = B_0 t {\hat {\bf z}} ~~ {\rm for} ~~ \rho <R ~~~~{\rm and}~~~~~
{\bf B} = {\bf 0} ~~ {\rm for} ~~ \rho \ge R ~,
\end{equation}
and the electric field connected with \eqref{sole-a} is 
\begin{equation}
\label{sole-e}
{\bf E} = - \frac{B_0 \rho}{2 c} {\hat {\boldsymbol \varphi}} ~~ {\rm for} ~~ \rho <R ~~~~{\rm and}~~~~~
{\bf E} = - \frac{B_0 R^2}{2 \rho c} {\hat {\boldsymbol \varphi}} ~~ {\rm for} ~~ \rho \ge R ~.
\end{equation}
The linear time dependence of the magnetic flux yields the above simple fields -- ${\bf A}, {\bf B}$ and ${\bf E}$. 
\footnote{If one considers sinusoidal time dependence the magnetic field will be non-zero outside the 
solenoid and the forms of both the electric and magnetic fields will involve Bessel and Neumann functions. \cite{macdougall}} 
The potential and fields in \eqref{sole-a} \eqref{sole-b} and \eqref{sole-e}
correspond to those given in \cite{griffiths} if the large time limit is taken. In \cite{griffiths} the linear increasing flux 
is turned on at $t=0$ whereas the potentials and fields above are linearly increasing for all time, but as $t$ 
becomes large the results of \cite{griffiths} yield those given above after units conversion.  

The 3-vector potential in \eqref{sole-a} can be ``redistributed" to form a new scalar and 3-vector potential, which gives the 
same electric and magnetic fields. The new 4-vector potential is
\begin{eqnarray}
\label{sole-a-2}
{A'} ^\mu &=& \left(\phi ', {\bf A}' \right) = \left( \frac{B_0 R^2 \varphi}{2 c} , 0, \left[ \frac{B_0 t \rho}{2} 
- \frac{B_0 R^2 t }{2 \rho} \right]{\hat {\boldsymbol \varphi}}, 0 \right) 
~~ {\rm for} ~~ \rho <R \\
~~~~{\rm and}~~~~~ \nonumber \\
\label{sole-a-2a}
{A'} ^\mu &=& \left(\phi ', {\bf A}' \right) = \left( \frac{B_0 R^2 \varphi}{2 c} , 0, 0, 0 \right) 
~~ {\rm for} ~~ \rho \ge R ~,
\end{eqnarray}
The two forms of the potentials for the time-dependent solenoid are related 
by a gauge transformation given by
\begin{equation}
\label{g-trans-sole}
{A'} ^\mu  \rightarrow A ^\mu + \partial ^\mu \chi ~~~{\rm with} ~~~ \chi = - \frac{B_0 R^2 \varphi t}{2}
\end{equation}
As in the previous time-independent case given in \eqref{a-time-1} \eqref{g-trans-gen}, ${\bf A}'$ has a
$\frac{1}{\rho}$ singularity and the gauge function, $\chi$, is non-single valued. 

We begin by first evaluating $\oint A_\mu dx^\mu$ for these two gauges as given in equations 
\eqref{sole-a} and \eqref{sole-a-2} \eqref{sole-a-2a}, on the closed spacetime path shown in figure \eqref{fig2}.

\subsection{Spacetime loop integral for the 4-vector potential from \eqref{sole-a}} 

\begin{figure}
  \centering
	\includegraphics[trim = 0mm 0mm 0mm 0mm, clip, width=15.0cm]{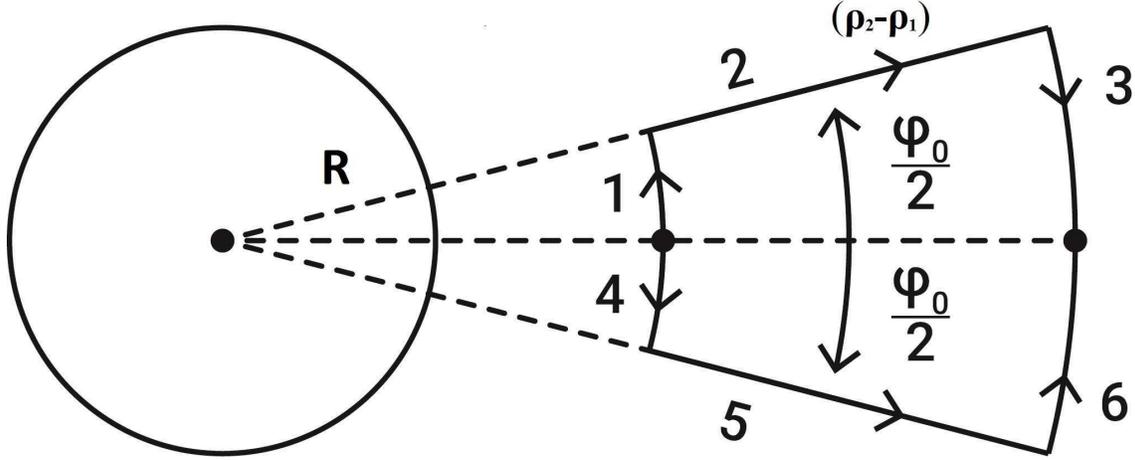}
\caption{{\scriptsize Closed loop path outside a solenoid with a time-changing magnetic flux. }}
\label{fig2}
\end{figure}

The evaluation of the loop integral $\oint A_\mu dx^\mu$ for the potential in \eqref{sole-a} is split into 
6 segments, as shown in Fig. \eqref{fig2}. Since in the end we want to connect our discussion of the time-dependent
Stokes' theorem with the time-dependent Aharonov-Bohm effect \cite{singleton}, we will take our paths to be those
traversed by a particle and thus the path lengths will be parameterized in terms of the particle's velocity. We will take
the linear speed of the particle to be the same along every path. In general Stokes' theorem could apply to paths 
which are not physically realizable paths for a particle. For example, even for the time-dependent case one could take a circular path 
around the solenoid at one instant in time. In this case one would just get $\oint {\bf A} \cdot d{\bf x} = \int {\bf B} \cdot d {\bf a}$
but the circular path would not be one that a real particle could travel. In addition, one would not touch on the
time-dependent nature of this case which, is one of the central points of this article. 

Stokes' theorem with time-independent fields has closed spatial loops when evaluating 
$\oint {\bf A} \cdot d{\bf x}$. Similarly, the time-dependent case has closed {\it spacetime} loops. 
To create a closed spacetime loop we consider two particles which begin 
at the same spacetime point, move apart and then come back together. We then time reverse 
the path of one of the particles and add this to the result of the path of the other particle. 
This is also the procedure for the Aharonov-Bohm effect with time-independent fields.  
The paths of the two moving particles are shown, with paths 1, 2 and 3 for one particle and paths 
4, 5 and 6 for the other particle. The arrows indicate the real directions of travel of each particle, 
starting from the middle of the inner arc at radius $\rho _1$ and ending at the middle of the outer arc at
radius $\rho_2$ . To obtain a closed spacetime path we apply the time reversal operator, $T[...]$, to the results of the 
line integrals for paths 4, 5 and 6. The operation of $T[...]$ takes $t \rightarrow -t$. The results of applying $T[...]$ 
to other physical quantities can be found in section 6.10 of reference. \cite{jackson} The oddness of ${\bf A}$ 
under time reversal ({\it i.e.} $T[{\bf A}] = -{\bf A}$) and the evenness of ${\bf x}$ under time reversal 
({\it i.e.} $T[{\bf x}] = {\bf x}$) implies that $T[\int {\bf A} \cdot d{\bf x} ] = - \int {\bf A} \cdot d{\bf x}$.
Thus the operation of $T[...]$ in segments 4, 5 and 6 has the effect of multiplying the results of these line 
integrals by $-1$. To get the closed spacetime loop we add the time reversed paths 4, 5 and 6 to the results from
paths 1, 2 and 3. Paths 1, 3, 4 and 6 cover an angle of $\varphi_0/2$, with 1 and 3 going between $0$ and $\varphi_0/2$, and 
with 4 and 6 going between $0$ and $-\varphi_0/2$.  The linear speed of the particles is taken to be constant throughout so that the angular 
speed along the inner paths, 1 and 4, is larger than the angular speed along the outer
paths, 3 and 6. The detailed definitions and calculations for each segment are given in Appendix I. 
Using these results we find that the closed spacetime loop is
\begin{eqnarray}
\label{path-tot-s}
\oint A_\mu dx^\mu &=& \left( \int _1  + \int _2  + \int _3 
+ T \left[ \int _6  + \int _5  + \int _4  \right] \right) {\bf A} \cdot d{\bf x} \nonumber \\
&=& - \frac{B_0 R^2 \varphi_0 ^2}{8 \omega} \left[ \frac{\rho_2}{\rho_1} +\frac{4 (\rho_2 -\rho_1)}{\rho_1 \varphi_0} + 1 \right] ~.
\end{eqnarray}
In \eqref{path-tot-s} $\omega$ is the angular speed at which paths 1 and 4 from Fig. \eqref{fig2} are traversed; $\rho _1$ is the distance from the center of
the solenoid to the inner paths 1 or 4;  $\rho _2$ is the distance from the center of the solenoid to the outer paths 3 and 6.
It is important to note for later that, when viewed from above, the direction of traversal of the
closed loop is clockwise. 

\subsection{Spacetime loop integral for the 4-vector potential from \eqref{sole-a-2} and \eqref{sole-a-2a}} 

The evaluation of the loop integral for the 4-potential in \eqref{sole-a-2}
now just involves the scalar potential $\oint A_\mu dx^\mu  \rightarrow -\oint \phi ' c dt$ since 
${\bf A}' =0$ outside the solenoid. The details of the calculation for the 6 segments can be found in 
Appendix II. Collecting together the results for the time-reversed paths 4, 5, and 6 (which, as in the 
previous case, changes the sign for these integrals as given in Appendix II) and adding these 
to the results from paths 1, 2 and 3 gives the closed spacetime loop result for the potential in this gauge as
\begin{eqnarray}
\label{path-tot-p}
\oint A_\mu ' dx^\mu &=& - \oint \phi ' c dt = - \left( \int _1  + \int _2  + \int _3   
+ T\left[ \int _4  + \int _5  + \int _6  \right] \right) \phi ' c dt \nonumber \\
&=& -  \frac{B_0 R^2 \varphi_0 ^2}{8 \omega} \left[ \frac{\rho_2}{\rho_1} +\frac{4 (\rho_2 -\rho_1)}{\rho_1 \varphi_0} + 1 \right] ~.
\end{eqnarray}
Comparing \eqref{path-tot-s} with \eqref{path-tot-p} we see that the two different gauges give the same result for $\oint A_\mu  dx^\mu$, 
as is expected for this gauge invariant quantity. We next calculate the spacetime area integral of the fields  
{\it i.e.} $\frac{1}{2} \int F_{\mu \nu} d \sigma ^{\mu \nu} = \int {\bf E} \cdot d{\bf x}~c dt + \int {\bf B} \cdot d{\bf a}$.

\subsection{Spacetime area integral for the fields from \eqref{sole-b} and \eqref{sole-e}} 

The evaluation of the spacetime area integral $\frac{1}{2} \int F_{\mu \nu} d \sigma ^{\mu \nu}$, for the fields 
in \eqref{sole-b} and \eqref{sole-e} on the spacetime area implied by Fig. \eqref{fig2}, reduces to $\int {\bf E} \cdot d{\bf x} c dt$
since ${\bf B} = 0$ outside the solenoid. The detailed calculations for $\int {\bf E} \cdot d{\bf x} c dt$
are given in Appendix III. 

We need to combine the results for the spacetime areas associated with the segments 1, 2, and 3 
with the time reversed spacetime areas associated with the time-reverse segments for 4, 5 and 6. From 
\cite{jackson}, ${\bf E}$ and ${\bf x}$ are even under time reversal ({\it i.e.} do not change sign)
whereas $t$ is odd ({\it i.e.} changes sign). Thus applying $T$ to $\int {\bf E} \cdot d{\bf x} ~ c dt$ changes the sign 
$T[ \int {\bf E} \cdot d{\bf x} ~ c dt] = - \int {\bf E} \cdot d{\bf x} ~ c dt$. 
This means that the time-reversed spacetime areas associated with segments 4, 5, and 6 
are equivalent to the spacetime areas associated with segments 1, 2, and 3. 
With all this in mind the total spacetime area integral is
\begin{eqnarray}
\label{e-area-tot-sol}
\int {\bf E} \cdot d{\bf x} ~ c dt &=&\left(\int _1 + \int _2 + \int _3 + T\left[\int _6 + \int _5 + \int _4 \right] \right) 
{\bf E} \cdot d{\bf x} ~ c dt \nonumber \\
&=& -  \frac{B_0 R^2 \varphi_0 ^2}{8 \omega} \left[ \frac{\rho_2}{\rho_1} +\frac{4 (\rho_2 -\rho_1)}{\rho_1 \varphi_0} + 1 \right]
\end{eqnarray}
We see that this result agrees with the spacetime line integral of the 4-vector potential
from \eqref{path-tot-s} or \eqref{path-tot-p}. Thus we find that, for this case, the time-dependent 4D version of 
Stokes' theorem is satisfied. In the next subsection we examine the case in which the path encloses the solenoid 
and therefore the spacetime area has a magnetic field contribution. 

\subsection{4D Stokes' Theorem for a path enclosing the solenoid} 

\begin{figure}
  \centering
	\includegraphics[trim = 0mm 0mm 0mm 0mm, clip, width=15.0cm]{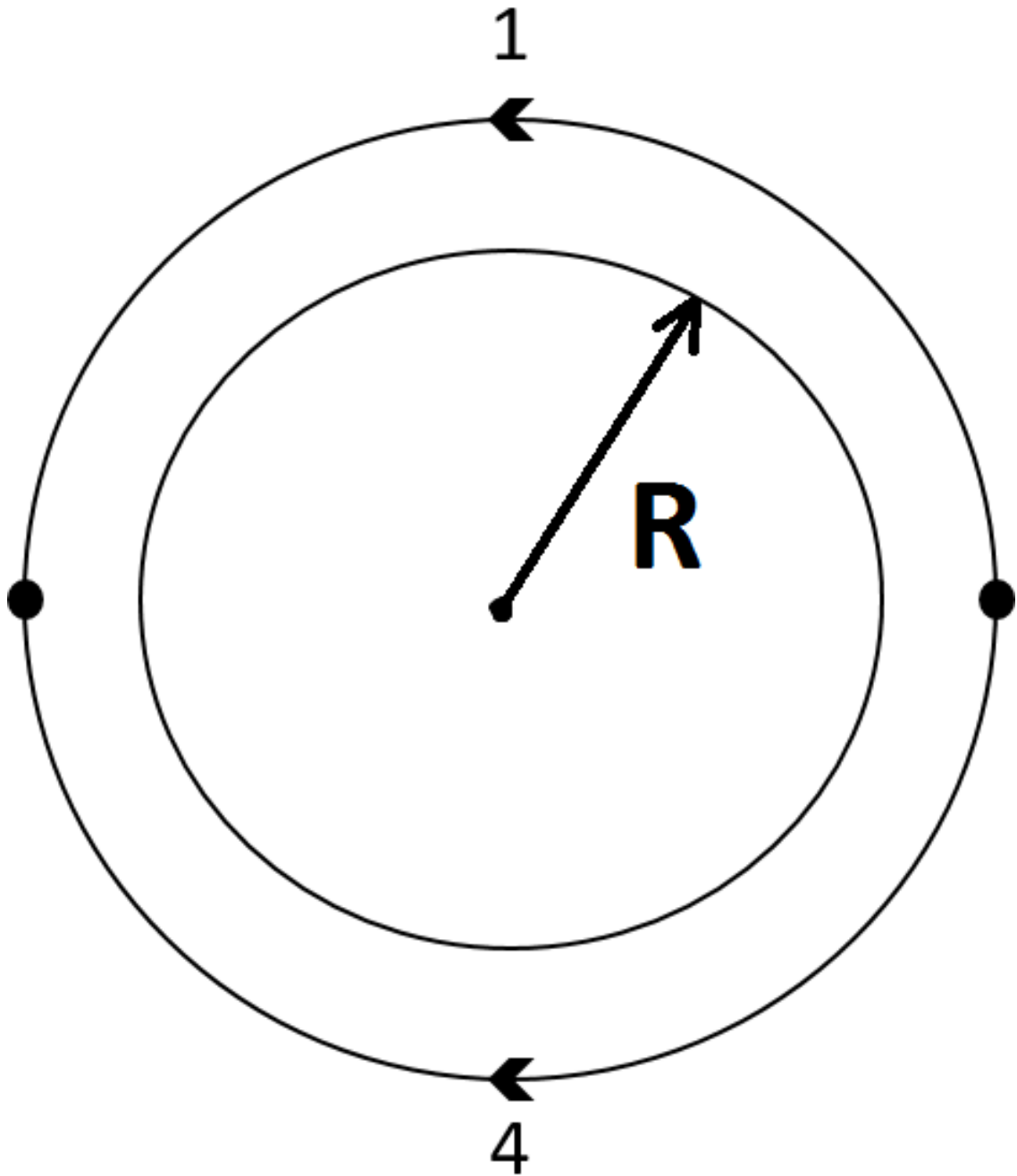}
\caption{{\scriptsize Closed loop path that encloses the solenoid with the time-changing magnetic flux. This closed loop is obtained
from the loop in figure \eqref{fig2} by discarding paths 2, 3, 5 and 6 and by extending paths 1 and 4 using $\varphi_0 = 2 \pi$.}}
\label{fig3}
\end{figure}

We now consider a closed spacetime loop that encloses the solenoid as shown in Fig. \eqref{fig3}. 
We will use the the results of the preceding subsections and the appendices to perform the calculations.
To enclose the solenoid with a spacetime path we eliminate paths 2, 3, 5 and 6 and then 
extend paths 1 and 4 around to form a closed loop. 

For the form of the 4-vector potential given in \eqref{sole-a}, the closed spacetime loop integral, $\oint A_\mu dx^\mu$, 
can be obtained using the line integrals of paths 1 and 4 (given by equations \eqref{path1s} and \eqref{path4s} respectively) with 
$\varphi_0 = 2 \pi$
\begin{equation}
\label{path14s}
\int _1 {\bf A} \cdot d{\bf x} = \frac{B_0 R^2 \pi ^2 }{4 \omega} ~~~{\rm and}~~~ \int _4 {\bf A} \cdot d{\bf x} = 
- \frac{B_0 R^2 \pi ^2 }{4 \omega} ~.
\end{equation} 
From \eqref{path14s}, one can obtain
\begin{equation}
\label{path-tot-14s}
\oint A_\mu dx^\mu = + \left( \int _1 {\bf A} \cdot d{\bf x} + T \int _4 {\bf A} \cdot d{\bf x}  \right) =  
\frac{B_0 R^2 \pi ^2 }{2 \omega} ~.
\end{equation} 
For this gauge $\oint A_\mu dx^\mu = -\oint \phi ~ c dt + \oint {\bf A} \cdot d{\bf x} \rightarrow  \oint {\bf A} \cdot d{\bf x}$
so only the 3-vector potential contributes. Also 
$\oint {\bf A} \cdot d{\bf x} = \int _1 {\bf A} \cdot d{\bf x} - \int _4 {\bf A} \cdot d{\bf x}$ 
since the direction of path 4 must be time-reversed to obtain a closed spacetime loop. It is worth noting 
that, when viewed from above, the path closes in a counterclockwise sense. This is the reverse 
of the closed path in Fig. \eqref{fig2}. This will have important consequences when we discuss the ``direction" 
of the spacetime area associated with the spacetime contours. 

\begin{figure}
  \centering
	\includegraphics[trim = 0mm 0mm 0mm 0mm, clip, width=15.0cm]{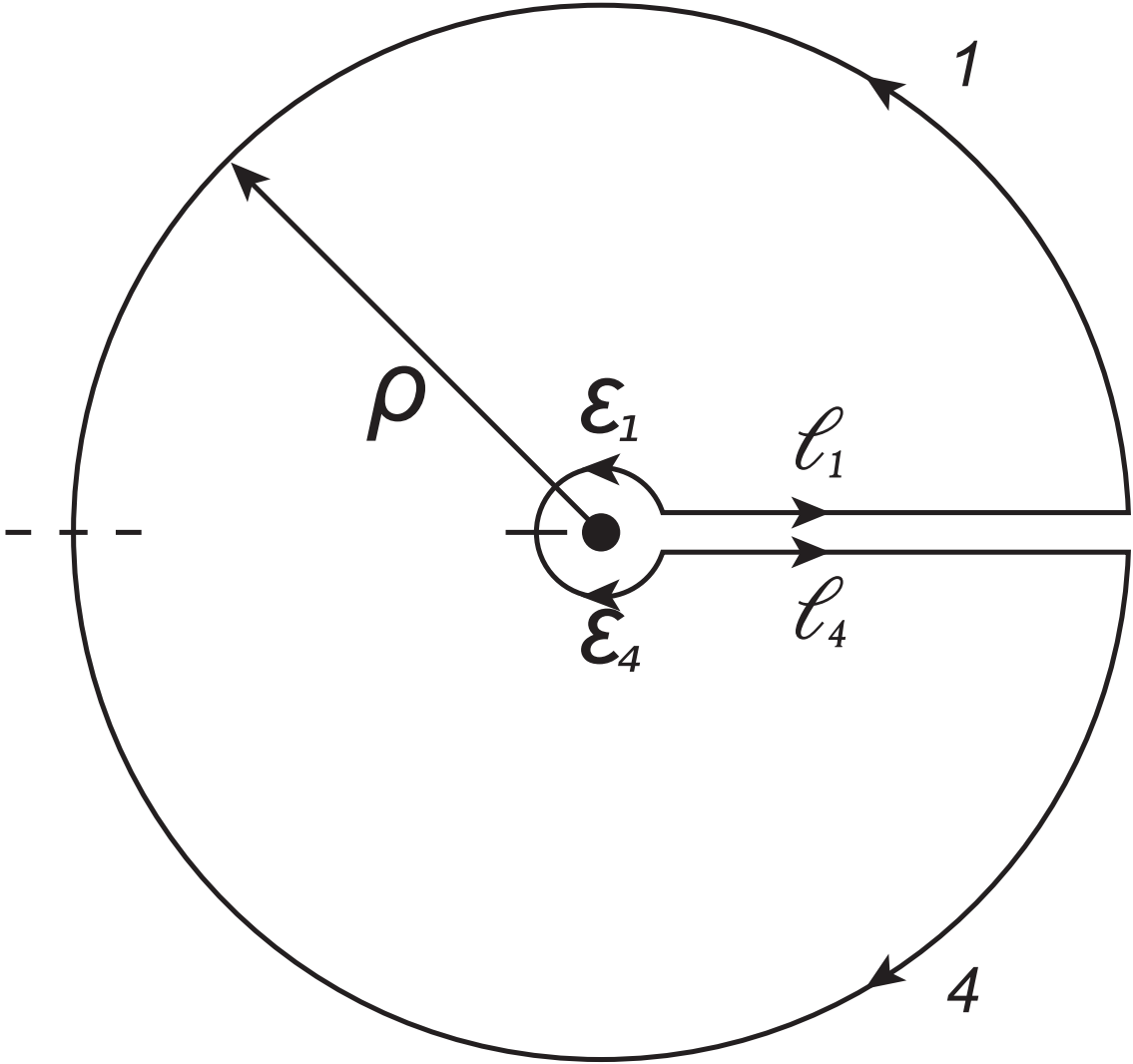}
\caption{{\scriptsize The spacetime loop for the case when the 4-vector potentials are given by \eqref{sole-a-2} and \eqref{sole-a-2a}.
The singularity at $\rho =0$ is excised by the small loop of radius $\epsilon \ll R$. The initial direction of integration of the different
paths are shown. To get a closed spacetime loop we time reverse path 4, path $l_4$ and path $\epsilon_1$. }}
\label{fig3a}
\end{figure}

Next we examine the form of the 4-vector potential given in \eqref{sole-a-2} and \eqref{sole-a-2a}, obtained from the form of the 
4-vector potential given in \eqref{sole-a} by the gauge transformation \eqref{g-trans-sole}. Due to the $\frac{1}{\rho}$
singularity for the $\rho <R$ 4-vector potential (see equation \eqref{sole-a-2}) the ``puncture" at $\rho =0$ needs to be removed 
using a contour similar to the one used in time-independent case (see Fig. \eqref{fig1}). The contour that we use now is shown in 
Fig. \eqref{fig3a}, with arrows indicating the direction of travel prior to time reversal of the paths 4, $l_4$ and $\epsilon_1$. 
We have contributions from the scalar potential only from paths 1 and 4 on the outer loop. The results from paths 1 and 4 
(given by \eqref{path1p} and \eqref{path4p} respectively) with  $\varphi_0 = 2 \pi$ are  
\begin{equation}
\label{path14p}
\int _1 \phi ' c dt =  \frac{B_0 R^2 \pi ^2}{4 \omega} ~~~~~ {\rm and} ~~~~~  
\int _4 \phi ' c dt =  - \frac{B_0 R^2 \pi ^2}{4 \omega} ~.
\end{equation}
Next we need to take into account the small interior circular path which we take to be of
radius $\epsilon \ll R$. Since $\phi ' = \frac{B_0 R^2 \varphi}{2 c}$ does not depend
on $\rho$, the results for the interior paths will be the same as those in \eqref{path14p} giving 
\begin{equation}
\label{path14pe}
\int _{\epsilon_1} \phi ' c dt =   \frac{B_0 R^2 \pi ^2}{4 \omega} ~~~~~ {\rm and} ~~~~~  
\int _{\epsilon_4} \phi ' c dt =   - \frac{B_0 R^2 \pi ^2}{4 \omega} ~.
\end{equation}
The subscripts $\epsilon _1 , \epsilon_4$ on the integrals above indicate that these are the interior half circle paths of 
radius $\epsilon \ll R$ corresponding to the outer paths 1 and 4 respectively. We have assumed the traversal of the inner
circle is at the same angular velocity, $\omega$, as the outer circle. We will see below that this is the only value for the angular
velocity that is able to excise the singularity.  

At this point we calculate the scalar potential contribution to $\oint A'_\mu dx^\mu$. From \eqref{path14p} \eqref{path14pe}, we get
\begin{eqnarray}
\label{oint-scalar}
\oint \phi ' c dt &=& \int _1 \phi ' c dt + \int _{\epsilon_4} \phi ' c dt  + \int _{l_1} \phi ' c dt 
+ T \left[ \int _4 \phi ' c dt  + \int _{\epsilon_1} \phi ' c dt + \int _{l_4} \phi ' c dt \right] \nonumber \\
&=& \frac{B_0 R^2 \pi ^2}{4 \omega} - \frac{B_0 R^2 \pi ^2}{4 \omega} + \int _{l_1} \phi ' c dt + \frac{B_0 R^2 \pi ^2}{4 \omega}
- \frac{B_0 R^2 \pi ^2}{4 \omega} - \int _{l_4} \phi ' c dt = 0
\end{eqnarray}
Although we did not explicitly calculate $\int _{l_1} \phi ' c dt$ and $\int _{l_4} \phi ' c dt$, it is clear that they are the 
same in magnitude and cancel after we apply time reversal. 

The only non-zero and uncanceled contribution to $\oint  A ' _\mu dx^\mu$ comes from the
interior 3-vector part of \eqref{sole-a-2}. The $-\frac{B_0 R^2 t}{2 \rho} {\hat {\boldsymbol \varphi}}$ piece gives the same result 
as \eqref{path14s} since the extra negative in this part of the 3-potential is balanced by the fact that the inner circular path is traversed 
in the opposite direction as the outer circle after time reversal. So we have
\begin{equation}
\label{path14e}
\int _{\epsilon_1} {\bf A}' \cdot d{\bf x} = \frac{B_0 R^2 \pi ^2 }{4 \omega} - \frac{B_0 \epsilon^2 \pi^2}{4 \omega}
~~~{\rm and}~~~ \int _{\epsilon_4} {\bf A}' \cdot d{\bf x} = 
- \frac{B_0 R^2 \pi ^2 }{4 \omega} + \frac{B_0 \epsilon^2 \pi^2}{4 \omega}~.
\end{equation} 
There is an additional contribution from the $\frac{B_0 \rho t}{2} {\hat {\boldsymbol \varphi}}$ term in the 3-potential relative to \eqref{path14s}, 
but in the limit $\epsilon \rightarrow 0$ this additional contribution is zero. From \eqref{path14e}, one can obtain
\begin{equation}
\label{path-tot-14e}
\oint A'_\mu dx^\mu = + \left( \int _{\epsilon_1} {\bf A}' \cdot d{\bf x} + T \int _{\epsilon_4} {\bf A}' \cdot d{\bf x}  \right) =  
\frac{B_0 R^2 \pi ^2 }{2 \omega} - \frac{B_0 \epsilon^2 \pi^2}{2 \omega} \rightarrow \frac{B_0 R^2 \pi ^2 }{2 \omega}~,
\end{equation}  
where at the end we have taken $\epsilon \rightarrow 0$. Thus by excising the singularity that exists in this gauge
we find that the results for the closed spacetime integrals given in \eqref{path-tot-14s} and \eqref{path-tot-14e} agree.  

We note that the interior circular path and the two paths in the ${\hat {\boldsymbol \rho}}$ direction are not physical paths that the 
particles traverse, but are simply artifacts used to excise the singularity at $\rho = 0$. Additionally, as we remarked above, 
we needed to arbitrarily take the traversal of the inner circle at the same angular velocity as the outer circle so that 
the singularity would be removed. This arbitrariness is absent from the time-independent case. In the time-dependent case, the ``strength"
of the $\frac{1}{\rho}$ singularity is $\propto \frac{B_0 R^2 t}{2}$, and thus changes linearly in time.     

\begin{figure}
 \centering
	\includegraphics[trim = 0mm 0mm 0mm 0mm, clip, width=15.0cm]{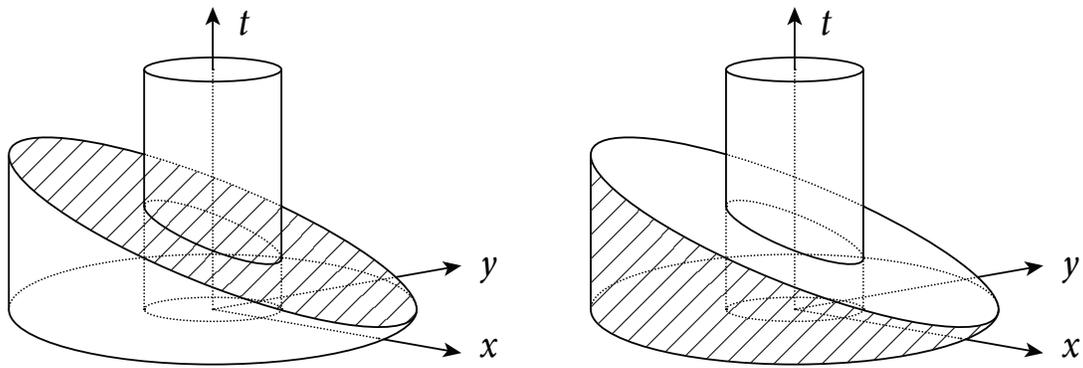}
\caption{{\scriptsize Two of the many possible spacetime areas associated with the spacetime loop, 
$\oint A_\mu dx^\mu$. The spacetime area on the left hand side ({\it i.e.} the slanted
top of the spacetime cylinder) gets a contribution only from the magnetic field. The spacetime area on the right hand side 
({\it i.e.} the side of the spacetime cylinder) gets a contribution only from the electric field.}}
\label{fig4}
\end{figure}

We now calculate the right hand side of Stokes' theorem -- the field version of the spacetime area integral,
$\frac{1}{2} \int F_{\mu \nu} d \sigma ^{\mu \nu} = \int {\bf E} \cdot d{\bf x}~c dt + \int {\bf B} \cdot d{\bf a}$. In this case,
we use the two spacetime surfaces shown in Fig. \eqref{fig4}, both of which span the spacetime path used in evaluating
$\oint A_\mu dx^\mu$. 

For the spacetime area that is the side of the spacetime cylinder -- the right side 
of Fig. \eqref{fig4} -- only the electric field will contribute 
{\it i.e.} $\frac{1}{2} \int F_{\mu \nu} d \sigma ^{\mu \nu} \rightarrow \int {\bf E} \cdot d{\bf x}~c dt$.
We can obtain this electric piece using our results from \eqref{e-field-a1} and \eqref{e-field-a4} with $\varphi_0 = 2 \pi$,
but with one subtlety: we need to reverse the signs relative to those given in \eqref{e-field-a1} and \eqref{e-field-a4} 
so that we have
\begin{equation}
\label{e-field-pi}
\int _1 {\bf E} \cdot d{\bf x} ~ c dt = + \frac{B_0 R^2 \pi ^2}{4 \omega} ~~~~;~~~~
\int _4 {\bf E} \cdot d{\bf x} ~ c dt = - \frac{B_0 R^2 \pi ^2}{4 \omega}.
\end{equation} 
The reason for the sign reversal is as follows: when we form a closed spacetime loop from the two contours in Fig. \eqref{fig3},
the direction of the spacetime loop is counterclockwise when the solenoid is viewed from above. As previously mentioned,
this is the reverse of the paths in Fig. \eqref{fig2}. For purely spatial
examples of Stokes' theorem, reversing the direction of traversal for the path reverses the direction of the 3-vector area
via the right-hand-rule. Here, although there is no equivalent right-hand-rule for the {\it spacetime} area, we nevertheless
need to reverse the spacetime area ``direction" when the closed spacetime path is traversed in the opposite direction. This
might be viewed as an extension of the right-hand-rule to spacetime paths and areas. One can
obtain this result on the ``direction" of the spacetime area less heuristically using differential forms and wedge 
product notation. Reviewing differential forms and wedge product notation is outside the scope of this article, but the 
interested reader can find nice expositions on this in the textbooks by Frankel \cite{frankel}, Felsager \cite{felsager}
and Ryder. \cite{ryder} To get the spacetime area contribution associated with the closed spacetime loop in Fig. \eqref{fig3}, we apply 
the time-reversal operator to area 4, which changes the sign of $\int \int _4 {\bf E} \cdot d{\bf x} ~ c dt$ since 
${\bf E}$ and $d {\bf x}$ are even under time reversal, but $dt$ is odd. We then add the time reversed result for
path 4 to the result for area 1 which gives 
\begin{equation}
\label{e-field-pi-a}
\int {\bf E} \cdot d{\bf x} dt = \left( \int _1  + T \int _4   \right) {\bf E} \cdot d{\bf x} ~ c dt
=  \frac{B_0 R^2 \pi ^2}{2 \omega}~.
\end{equation} 
This result agrees with the results for the loop integral for the 4-vector potentials in the two different 
gauges, given in \eqref{path-tot-14s} and \eqref{path-tot-14e}.

We now calculate $\frac{1}{2} \int F_{\mu \nu} d \sigma ^{\mu \nu}$ for the spacetime area on the left hand side
of Fig. \eqref{fig4} {\it i.e.} the slanted ``top" of the spacetime cylinder. For this spacetime area, only the
magnetic field contributes {\it i.e.} $\frac{1}{2} \int F_{\mu \nu} d \sigma ^{\mu \nu} \rightarrow \int {\bf B} \cdot d{\bf a}$.
For the previous case, shown in Fig. \eqref{fig2}, our spacetime loop enclosed a region where ${\bf B} =0$. 
Now from the right hand side of Fig. \eqref{fig4} we can see that the spacetime area will have ${\bf B} \ne 0$ so we expect 
$\int {\bf B} \cdot d{\bf a} \ne 0$. For the spacetime area associated with path 1 we have 
\begin{equation}
\label{b-field-pi-1}
\int _0 ^\pi \int _0 ^R {\bf B} \cdot {\hat {\bf z}} \rho d \rho d \varphi = \frac{B_0 R^2}{2 \omega} \int _0 ^\pi \varphi d \varphi 
= \frac{B_0 R^2 \pi^2 }{4 \omega}  ~,
\end{equation} 
where we have used $t= \varphi / \omega$ to write the magnetic field magnitude as $B_0 t \rightarrow B_0 \varphi / \omega$. 
Note that here the area vector points along the positive z-axis $d{\bf a} = {\hat {\bf z}} \rho d \rho d \varphi$. 
For the spacetime area associated with segment 4 we have
\begin{equation}
\label{b-field-pi-4}
\int _0 ^{-\pi} \int _0 ^R {\bf B} \cdot (-{\hat {\bf z}}) \rho d \rho d \varphi = - \frac{B_0 R^2}{2 \omega} \int _0 ^{-\pi} 
\varphi d\varphi = - \frac{B_0 R^2 \pi^2 }{4 \omega}  ~.
\end{equation} 
Again we have used $t= \varphi / \omega$ and $d\varphi = \omega dt$. Here the area vector points along the negative z-axis 
($d{\bf a} = - {\hat {\bf z}} \rho d \rho d \varphi$) since the path is traversed in a clockwise direction. To get
the spacetime area associated with the closed spacetime loop, we add the result of \eqref{b-field-pi-1} to the time
reversed result of \eqref{b-field-pi-4}
\begin{equation}
\label{b-field-pi-area}
\int {\bf B} \cdot d{\bf a} = \int_1 {\bf B} \cdot d{\bf a} + T \left[\int_4 {\bf B} \cdot d{\bf a} \right] =
\int_1 {\bf B} \cdot d{\bf a} - \int_4 {\bf B} \cdot d{\bf a} = \frac{B_0 R^2 \pi^2 }{2 \omega}  ~.
\end{equation} 
This result agrees with the results for the loop integrals for the 4-vector potentials in the two different 
gauges given in \eqref{path-tot-14s} and \eqref{path-tot-14e}, and agrees with the result for the spacetime area
integral from \eqref{e-field-pi-a}. This is reminiscent of the charging capacitor demonstration of Maxwell's
displacement current. \cite{d-current} In this example, there is a circular loop that encloses a wire which is charging up a capacitor. 
The closed loop integral of the magnetic field ({\it i.e.} $\oint {\bf B} \cdot d{\bf x}$) gives
${\bf B} \propto \frac{I(t)}{\rho} {\hat {\boldsymbol \varphi}}$.
If the surface chosen to span this loop cuts through the wire then $I(t)$ is the current due to charges. However, 
the surface can be chosen so that it goes between the capacitor plates where one has no current from charges. In this 
region one does have a time-changing electric field and so the contribution to the magnetic field
loop integral now comes from the displacement current. Here, the loop 
integral of the vector potential is related to the spacetime area integral of the electric field 
in one case, and related to the spacetime area integral of the magnetic field in the other case.   

\section{Summary and Conclusions}
In this work, we have given explicit examples of Stokes' theorem in the case where there exists time-dependent fields 
({\it i.e.} 4D Stokes' theorem). There are many examples of Stokes' theorem as it applies to time-independent fields 
({\it i.e.} 3D Stokes' theorem), but these are perhaps the first explicit worked out examples of 
Stokes' theorem with time-dependent fields. In fact, there are some claims that Stokes' theorem does not apply to 
time-dependent fields (see the first footnote on page 305 and the first paragraph on page 312 of reference \cite{brown}). 
Despite these assertions, we have found that Stokes' theorem \emph{can} be applied to time-dependent cases if care is taken in 
how the spacetime loop is closed. In particular we investigated an infinite solenoid with a linearly increasing magnetic flux. 
There we showed $\oint A _\mu dx^\mu = \frac{1}{2} \int F_{\mu \nu} d\sigma ^{\mu \nu}$ for the spacetime loop closed 
outside the solenoid -- see Fig. \eqref{fig2}. We also showed that $\oint A' _\mu dx^\mu$, for the closed spacetime loop from
Fig. \eqref{fig2}, remained the same for a gauge transformed 4-potential as given
in \eqref{sole-a-2} \eqref{sole-a-2a}, demonstrating the gauge invariance of $\oint A _\mu dx^\mu$.

We also investigated the more subtle case when the spacetime loop enclosed the solenoid. Here we also calculated 
$\oint A_\mu dx^\mu$ in the two gauges given in \eqref{sole-a} and \eqref{sole-a-2} \eqref{sole-a-2a}. For the 4-vector
potential (obtained with the multi-valued gauge function $\chi$)
given in \eqref{sole-a-2} \eqref{sole-a-2a}, we had to excise the singularity 
at $\rho =0$ via the contour given in Fig. \eqref{fig3a}. In terms of the spacetime area integral of
the fields we carried out the calculation using the two spacetime areas
given in Fig. \eqref{fig4}: (i) the side of the spacetime cylinder where only the electric field 
contributed; (ii) the slanted top of the spacetime cylinder where only the magnetic field contributed. 
The result of $\frac{1}{2}\int F_{\mu \nu} d \sigma ^{\mu \nu}$ came either from the side spacetime area
or the slanted top spacetime area. This is similar to the standard example used to demonstrate Maxwell's 
displacement current.   

The time-dependent Stokes' theorem is closely related to the time dependent Aharonov-Bohm effect and the
the discussion in this article has been closely guided by this connection. The spacetime contours from Figs.
\eqref{fig2} \eqref{fig3} were taken to be those that a real particle could traverse since this is what
occurs in the Aharonov-Bohm effect. Although much work has been done on the time-independent
Aharonov-Bohm effect, much less has been done in regard to the time-dependent Aharonov-Bohm effect. There are
only two experiments that we have found which have been done on the time-dependent Aharonov-Bohm effect --
one accidental experiment \cite{marton} and one purposeful experiment. \cite{ageev} Two theoretical papers 
\cite{werner, macdougall2} were written in an attempt to explain the surprising non-result of the accidental 
experiment of Marton {\it et al.} \cite{marton} More recently, there have been some theoretical papers
\cite{chiao, singleton, bright} dealing with the time-dependent Aharonov-Bohm effect. Still,
aside from the two experiments in \cite{marton, ageev} (both of which gave unclear results),
there is little in the way of experimental results for the time-dependent Aharonov-Bohm effect. 

As a final comment we note that the electric field associated with the solenoid, given in \eqref{sole-e}, is of the form
one would expect for a current of magnetic charges -- just as a current of electric charges produces a magnetic field 
of the form ${\bf B} \propto \frac{1}{\rho} {\hat {\boldsymbol \varphi}}$, so too a current of magnetic charges would produce an 
electric field of the form ${\bf E} \propto \frac{1}{\rho} {\hat {\boldsymbol \varphi}}$. This connection to magnetic charge also
offers another example of a non-single valued gauge transformation. The following 3-vector potential 
(we now use spherical polar coordinates $r, \theta, \varphi$, rather than the cylindrical coordinates, $\rho, \varphi, z$) 
yields a monopole magnetic field ${\bf B} = \nabla \times {\bf A} = g {\hat {\bf r}}/ r^2$
\begin{equation}
\label{monopole}
{\bf A} _{{\rm monopole}} = \frac{g (1-\cos \theta )}{r\sin \theta} {\hat {\boldsymbol \varphi}} ~.
\end{equation} 
The vector potential in \eqref{monopole} is single valued, but it has the usual Dirac string singularity pathology 
along the negative z-axis {\it i.e.} $\theta = \pi$. One can also obtain a magnetic monopole field
from ${\bf A} _{monopole} = - \frac{g (1+\cos \theta )}{r\sin \theta} {\hat {\boldsymbol \varphi}}$ which has a
Dirac string singularity along the positive z-axis {\it i.e.} $\theta = 0$. These two forms of the monopole
3-vector potential are related by the gauge transformation ${\bf A} \rightarrow {\bf A} - \nabla \chi$
with $\chi = 2 g \varphi$. In this case the gauge function $\chi$ is non-single valued, but 
the two forms of the gauge potential ${\bf A} _{{\rm monopole}}$ {\it are} single valued. Thus, this is 
not exactly like the 4-potentials and gauge transformation for the time-dependent solenoid case,
given in \eqref{sole-a-2} \eqref{sole-a-2a} and \eqref{g-trans-sole},
where both the potentials and gauge function were non-single valued. 

It is easy to see that one can also get a magnetic monopole  field from the following alternative 3-vector 
potential \cite{arfken, pal, macdougall}
\begin{equation}
\label{monopole1}
{\bf A} _{{\rm monopole}} = - \frac{g \varphi \sin \theta }{r} {\hat {\boldsymbol \theta}} ~,
\end{equation} 
which does not have the Dirac string singularity of \eqref{monopole}, but is non-single valued 
due to the $\varphi$ dependence of $A_\theta$. The two vector potentials in 
\eqref{monopole} and \eqref{monopole1} are related by a gauge transformation of the form
\begin{equation}
\label{mono-gauge}
A^\mu \rightarrow A^\mu + \partial ^\mu \chi ~~~;~~~ \chi = - g (1 - \cos \theta) \varphi
\end{equation}
Here we see that both the gauge transformation function $\chi$ in \eqref{mono-gauge} {\it and} the
3-vector gauge potential \eqref{monopole1} are non-single valued, which then is similar to the situation
for the time-dependent solenoid as given in \eqref{sole-a-2} , \eqref{sole-a-2a}, \eqref{g-trans-sole}. \\

{\par\noindent {\bf Acknowledgments:}} DS is supported by a 2015-2016 Fulbright Scholars Grant to Brazil
and by grant $\Phi.0755$  in fundamental research in natural sciences by the Ministry of 
Education and Science of Kazakhstan. DS wishes to thank the ICTP-SAIFR in S{\~a}o Paulo for it hospitality. 
We gratefully acknowledge Joe Deutscher for help with the figures in the paper. \\

\appendix{{\bf Appendix I}} \\

In this appendix we carry out the details of the 6 line integrals for $\oint A_\mu dx^\mu$ with the 4-vector potential
from \eqref{sole-a}. For path 1, the vector potential from \eqref{sole-a} is 
${\bf A} = \frac{B_0 t R^2}{2 \rho_1} {\hat {\boldsymbol \varphi}}$ 
since the radius on this path is $\rho_1$. The infinitesimal path length element is 
$d{\bf x} = \rho _1 d \varphi {\hat {\boldsymbol \varphi}}$. 
The path is traversed at a constant angular speed 
$\omega$ so that we have the relationship $\varphi = \omega t \rightarrow d \varphi = \omega dt$. 
The particle starts at $\varphi =0$ and $t=0$ and ends at $\varphi = \frac{\varphi_0}{2}$ at $t= \frac{\varphi_0}{2 \omega}$, 
yielding
\begin{equation}
\label{path1s}
\int _1 {\bf A} \cdot d{\bf x} = \frac{B_0 R^2}{2} \int _0 ^{\varphi_0 / 2} t d \varphi = 
\frac{B_0 R^2 \omega}{2} \int _0 ^{\varphi_0 / 2 \omega} t dt = \frac{B_0 R^2 \varphi_0 ^2}{16 \omega}
\end{equation}
For path 2 we have ${\bf A} = \frac{B_0 t R^2}{2 \rho} {\hat {\boldsymbol \varphi}}$ 
(now $\rho$ varies) and $d{\bf x} = d \rho {\hat {\boldsymbol \rho}}$. Thus
${\bf A} \cdot d{\bf x} =0$ and we get no contribution to the loop integral from this line segment.
We have taken the velocity along path 2 to be the same as the velocity along path 1 (namely $\rho_1 \omega$) and the distance 
traveled is $\rho_2 - \rho_1$ (the distance from line segment 1 to line segment 3). Thus for path 2 we have 
\begin{equation}
\label{path2s}
\int _2 {\bf A} \cdot d{\bf x} \propto \int_2  {\hat {\boldsymbol \varphi}} \cdot {\hat {\boldsymbol \rho}} d \rho = 0 ~.
\end{equation}
Although the line integral for path 2 is zero, time passes in the traversal of the path. This will have an effect on the 
result for line segment 3. The amount of time that passes during the traversal of path 2 is 
$\Delta t _2 = \frac{\rho_2 - \rho_1}{\rho_1 \omega}$.  
For path 3 from \eqref{sole-a} we have ${\bf A} = \frac{B_0 t R^2}{2 \rho_2} {\hat {\boldsymbol \varphi}}$, 
since the radius on this path is $\rho_2$.
The infinitesimal length element is $d{\bf x} = - \rho _2 d \varphi {\hat {\boldsymbol \varphi}}$. 
The path is traversed at a constant angular speed
$\omega' = \omega \frac{\rho_1}{\rho_2}$ (the angular speed is slower) but this ensures that the linear speed along paths 1, 2 and 3 
is the same \footnote{Here the requirement that the linear speed be the same along all line segments is a convenience. However,
since one of the applications of our analysis is to the Aharonov-Bohm effect where one wants to eliminate or minimize the
external forces on the particle tracing out the spacetime path we take the speed to be constant Of course at the bends in the paths
there will be forces but these can be thought of as the bending forces due to crystalline diffraction such as in the experiment in
\cite{marton}}. We now have the relationship $\varphi = \omega' t \rightarrow d \varphi = \omega' dt$. The particle starts at 
$\varphi = \frac{\varphi_0}{2} $ and $t_i= \frac{\varphi_0}{2 \omega} + \frac{\rho_2 - \rho_1}{\rho_1 \omega}$ (this offset time 
$\frac{\rho_2 - \rho_1}{\rho_1 \omega}$ is connected with the traversal of path 2). The particle ends at $\varphi = 0$ at 
$t_f= t_i + \frac{\varphi_0}{2 \omega'}$, yielding
\begin{eqnarray}
\label{path3s}
\int _3 {\bf A} \cdot d{\bf x} &=& -\frac{B_0 R^2}{2} \int _{t_i} ^{t_f} t d \varphi = 
-\frac{B_0 R^2 \omega'}{2} \int _{t_i} ^{t_f} t dt  = -\frac{B_0 R^2 \omega'}{4} t^2 {\bigg\vert}_{t_i} ^{t_f} \nonumber \\
&=& -\frac{B_0 R^2 \varphi_0 ^2}{16 \omega} \left[ \frac{\rho_2}{\rho_1} +\frac{4 (\rho_2 -\rho_1)}{\rho_1 \varphi_0} +2 \right]
\end{eqnarray}

Next we calculate $\int {\bf A} \cdot d{\bf x}$ for line segments 4, 5 and 6. The integral $\int _4 {\bf A} \cdot d{\bf x}$ 
is the same as $\int _1 {\bf A} \cdot d{\bf x}$ except $d{\bf x} = - \rho _1 d \varphi {\hat {\boldsymbol \varphi}}$ which changes the
final result by a sign
\begin{equation}
\label{path4s}
\int _4 {\bf A} \cdot d{\bf x} =  - \frac{B_0 R^2 \varphi_0 ^2}{16 \omega}
\end{equation}
The integral $\int _5 {\bf A} \cdot d{\bf x}$, like $\int _2 {\bf A} \cdot d{\bf x}$ is zero since 
\begin{equation}
\label{path5s}
\int _5 {\bf A} \cdot d{\bf x} \propto \int_5  {\hat {\boldsymbol \varphi}} \cdot {\hat {\boldsymbol \rho}} d \rho = 0 ~.
\end{equation} 
Again time $\Delta t _5 = \frac{\rho_2 - \rho_1}{\rho_1 \omega}$ passes during the traversal of path 5,
affecting the result of path 6. Finally, path 6 is similar to path 3 except 
$d{\bf x} =  \rho _2 d \varphi {\hat {\boldsymbol \varphi}}$ which changes the final overall sign. Similar to path 3, path 6 is traversed at a 
constant angular speed $\omega' = \omega \frac{\rho_1}{\rho_2}$ which ensures that the linear speed along path 
6 and path 4 are the same (in fact the speed along all six segments is taken to be the same). As before, we have 
the relationship $\varphi = \omega' t \rightarrow d \varphi = \omega' dt$. The particle starts at 
$\varphi = - \frac{\varphi_0}{2} $ and $t_i = \frac{\varphi_0}{2 \omega} + \frac{\rho_2 - \rho_1}{\rho_1 \omega}$ and 
ends at $\varphi = 0$ and $t_f= t_i + \frac{\varphi_0}{2 \omega'}$, yielding
\begin{equation}
\label{path6s}
\int _6 {\bf A} \cdot d{\bf x}
= \frac{B_0 R^2 \varphi_0 ^2}{16 \omega} \left[ \frac{\rho_2}{\rho_1} +\frac{4 (\rho_2 -\rho_1)}{\rho_1 \varphi_0} +2 \right]
\end{equation}
Notice that the result for path 6 is equivalent to the negative of path 3. \\  

\appendix{{\bf Appendix II}} \\

In this appendix we carry out the details of the 6 line segment integrals for $\oint A_\mu dx^\mu$ for the 4-vector potential
from \eqref{sole-a-2}. For  path 1 we have
\begin{equation}
\label{path1p}
\int _1 \phi ' c dt =  \frac{B_0 R^2}{2} \int _0 ^{\varphi_0 / 2 \omega} \varphi dt = 
\frac{B_0 R^2 \omega}{2} \int _0 ^{\varphi_0 / 2 \omega} t dt =  \frac{B_0 R^2 \varphi_0 ^2}{16 \omega} ~,
\end{equation}
where we use $\varphi = \omega t$. For path 2, the scalar potential is constant $\phi ' =\frac{B_0 R^2 \varphi_0}{4}$
and the time to traverse path 2 is, as in Appendix I, $\Delta t = \frac{\rho_2 - \rho_1}{\rho_1 \omega}$. So
for path 2 we have
\begin{equation}
\label{path2p}
\int _2 \phi ' c dt =  \frac{B_0 R^2 \varphi _0}{4} \int _0 ^{\Delta t} dt = 
 \frac{B_0 R^2 \varphi _0}{4} \left( \frac{\rho_2 - \rho_1}{\rho_1 \omega} \right)~.
\end{equation}
For path 3 we use the relationship $\varphi = \frac{\varphi_0}{2} - \omega ' t$. The integral is
\begin{equation}
\label{path3p}
\int _3 \phi ' c dt =  \frac{B_0 R^2 }{2} \int _0 ^{\varphi_0 / 2 \omega '} \varphi dt = 
 \frac{B_0 R^2 }{2} \left( \frac{\varphi _0}{2} t - \frac{1}{2} \omega ' t^2 \right) {\bigg\vert}_0 ^{\varphi_0 / 2 \omega '}
=  \frac{B_0 R^2 \varphi_0 ^2 \rho _2}{16 \omega \rho_1} .
\end{equation}

Next we calculate $\int \phi ' c dt$ for line segments 4, 5, and 6. Line segment 4 is similar to path 1 except
that $\varphi = - \omega t$ which then changes the sign of the result
\begin{equation}
\label{path4p}
\int _4 \phi ' c dt = -\frac{B_0 R^2 \varphi_0 ^2}{16 \omega} ~.
\end{equation}
Path 5 is similar to path 2 except now the scalar potential takes a different constant value
$\phi ' = - \frac{B_0 R^2 \varphi_0}{4}$.
\begin{equation}
\label{path5p}
\int _5 \phi ' c dt = 
- \frac{B_0 R^2 \varphi _0}{4} \left( \frac{\rho_2 - \rho_1}{\rho_1 \omega} \right)~.
\end{equation}
Path 6 is similar to path 3 except now we have $\varphi = - \frac{\varphi_0}{2} + \omega ' t$. The integral is
\begin{equation}
\label{path6p}
\int _6 \phi ' c dt =   - \frac{B_0 R^2 \varphi_0 ^2 \rho _2}{16 \omega \rho_1} . \\
\end{equation}

\appendix{{\bf Appendix III}} \\

In this appendix we carry out details of the spacetime area integral of the electric and magnetic fields for the
spacetime loop given in Fig. \ref{fig2} 

For the spacetime area connected with path 1, the
electric field and infinitesimal line element are ${\bf E} = -\frac{B_0 R^2}{2 c \rho _1} {\hat {\boldsymbol \varphi}}$ and 
$d {\bf x} = \rho_1 d \varphi {\hat {\boldsymbol \varphi}}$ respectively, so 
\begin{equation}
\label{e-field-a1}
\int _1 {\bf E} \cdot d{\bf x} ~ c dt = - \int _0 ^{\varphi_0 / 2 \omega} dt \int _0 ^{\omega t} \frac{B_0 R^2}{2} d \varphi
= - \int _0 ^{\varphi_0 / 2 \omega} \left( \frac{B_0 R^2 \omega t}{2} \right) dt = - \frac{B_0 R^2 \varphi_0 ^2}{16 \omega} ~.
\end{equation} 
For the spacetime area connected with path 2, the electric field is ${\bf E} = -\frac{B_0 R^2}{2 c \rho } {\hat {\boldsymbol \varphi}}$ --
now the radial coordinate $\rho$ is not fixed at $\rho = \rho_1$ but rather runs from $\rho_1$ to $\rho_2$. The infinitesimal
line element is \footnote{Note that the spacetime area
integration connected with path 2 runs along $d\varphi$. It is not along the linear path 2 which would be an integration
along $d \rho$.} $d {\bf x} = \rho d \varphi {\hat {\boldsymbol \varphi}}$ so that for each $\rho$ along path 2 (with $\rho_1 \le \rho \le \rho_2$) 
the $d \varphi$ integration runs from $\varphi =0$ to $\varphi = \varphi_0 /2$. The integration over $dt$ runs from $t=0$ to 
$t=\frac{\rho_2-\rho_1}{\rho_1 \omega}$ which corresponds to moving from $\rho_1$ to $\rho_2$ at a speed of $\rho_1 \omega$. The 
$dt$ integration in combination with the $d \varphi$ integration sweeps out the spacetime area. Thus for path 2 the spacetime area integral is
\begin{eqnarray}
\label{e-field-a2}
\int _2 {\bf E} \cdot d{\bf x} ~ c dt &=& - \int _0 ^{\frac{\rho_2-\rho_1}{\rho_1 \omega}} dt 
\int _0 ^{\varphi_0/2} \frac{B_0 R^2}{2} d \varphi \nonumber \\
&=& - \int _0 ^{\frac{\rho_2-\rho_1}{\rho_1 \omega}} \left( \frac{B_0 R^2 \varphi_0}{4} \right) dt = 
- \frac{B_0 R^2 \varphi_0 ^2}{4 \rho_1 \omega} (\rho_2 - \rho_1 )~.
\end{eqnarray}
For the spacetime area connected with path 3 the electric field is ${\bf E} = -\frac{B_0 R^2}{2 c \rho_2 } {\hat {\boldsymbol \varphi}}$
and the infinitesimal line element is $d {\bf x} = \rho_2 d \varphi {\hat {\boldsymbol \varphi}}$. The $d \varphi$ integration
goes from $\varphi=0$ to $\varphi = \frac{\varphi_0}{2} - \omega ' t$ (the angular velocity is
$\omega ' = \frac{\rho_1}{\rho_2} \omega$ so that the linear speed is the same on each path). The $dt$ integration runs from $t=0$ to 
$t= \varphi_0 /2 \omega'$. Thus the spacetime integral connected with path 3 is
\begin{eqnarray}
\label{e-field-a3}
\int _3 {\bf E} \cdot d{\bf x} ~ c dt &=& - \int _0 ^{\varphi_0 / 2 \omega'} dt 
\int _0 ^{\frac{\varphi_0}{2} - \omega ' t} \frac{B_0 R^2}{2} d \varphi \nonumber \\
&=& - \int _0 ^{\varphi_0 / 2 \omega'} \frac{B_0 R^2 }{2} \left(  \frac{\varphi_0}{2} - \omega ' t \right) dt = 
- \frac{B_0 R^2 \varphi_0 ^2 \rho_2}{16 \omega \rho_1} ~.
\end{eqnarray} 
The spacetime area integration connected with path 4 is similar to the integration connected with path 1, except here
the limits on the $d \varphi$ integration run from $\varphi=0$ to $\varphi = -\omega t$, yielding  
\begin{equation}
\label{e-field-a4}
\int _4 {\bf E} \cdot d{\bf x} c dt = - \int _0 ^{\varphi_0 / 2 \omega} dt \int _0 ^{-\omega t} \frac{B_0 R^2}{2} d \varphi
= \int _0 ^{\varphi_0 / 2 \omega} \left( \frac{B_0 R^2 \omega t}{2} \right) dt =  \frac{B_0 R^2 \varphi_0 ^2}{16 \omega} ~.
\end{equation} 
Note by comparing \eqref{e-field-a1} and \eqref{e-field-a4} one sees
$\int _4 {\bf E} \cdot d{\bf x} c dt = - \int _1 {\bf E} \cdot d{\bf x} c dt$. The spacetime area 
integration connected with path 5 is similar to the integration connected with path 2, except the limits
on the $d \varphi$ integration run from $\varphi=0$ to $\varphi = -\varphi_0/2$, yielding  
\begin{eqnarray}
\label{e-field-a5}
\int _5 {\bf E} \cdot d{\bf x} ~ c dt &=& - \int _0 ^{\frac{\rho_2-\rho_1}{\rho_1 \omega}} dt 
\int _0 ^{-\varphi_0/2} \frac{B_0 R^2}{2} d \varphi \nonumber \\
&=& \int _0 ^{\frac{\rho_2-\rho_1}{\rho_1 \omega}} \left( \frac{B_0 R^2 \varphi_0}{4} \right) dt = 
\frac{B_0 R^2 \varphi_0 ^2}{4 \rho_1 \omega} (\rho_2 - \rho_1 )~.
\end{eqnarray}
Note that comparing \eqref{e-field-a2} and \eqref{e-field-a5} illustrated that 
$\int _5 {\bf E} \cdot d{\bf x} c dt = - \int _2 {\bf E} \cdot d{\bf x} c dt$.
Finally, the spacetime area integration connected with path 6 is similar to the integration connected with
path 3, except here the limits on the $d \varphi$ integration run from $\varphi=0$ to $\varphi = \omega' t -\varphi_0/2$. 
\begin{eqnarray}
\label{e-field-a6}
\int _6 {\bf E} \cdot d{\bf x} ~ c dt &=& - \int _0 ^{\varphi_0 / 2 \omega'} dt 
\int _0 ^{\omega ' t - \frac{\varphi_0}{2}} \frac{B_0 R^2}{2} d \varphi \nonumber \\
&=& - \int _0 ^{\varphi_0 / 2 \omega'} \frac{B_0 R^2 }{2} \left( \omega ' t - \frac{\varphi_0}{2} \right) dt = 
\frac{B_0 R^2 \varphi_0 ^2 \rho_2}{16 \omega \rho_1} ~.
\end{eqnarray} 
Again note that comparing \eqref{e-field-a3} and \eqref{e-field-a6} illustrates that 
$\int _6 {\bf E} \cdot d{\bf x} c dt = - \int _3 {\bf E} \cdot d{\bf x} c dt$.

As a final comment, the ``direction" of the spacetime area associated with the closed spacetime 
paths of Fig. \eqref{fig2} or Fig. \eqref{fig3} does not have a well known right hand rule as is the case for purely
spatial contours and areas. The directionality that we have chosen above for the spacetime area is taken so that
the area integral of the fields in \eqref{e-area-tot-sol} agrees with the results of the contour integrals
of the two vector potentials given in \eqref{path-tot-s} and \eqref{path-tot-p} for the closed 
contour given in Fig. \eqref{fig2} ({\it i.e.} this choice ensures that Stokes' theorem works out for Fig. 
\eqref{fig2}). In contrast, for the path which encloses the solenoid in Fig. \eqref{fig3} the direction in which the
path is traversed, and thus the spacetime ``area" direction, is reversed relative to the path in Fig \eqref{fig2}.
Thus the spacetime area associated with the contour in Fig. \eqref{fig3} must have the opposite sign from
the spacetime area which comes from the contour in Fig. \eqref{fig2}. This again ensures that Stokes' theorem 
works out for the contour which encloses the solenoid. One can regard the procedure described above, with the determining 
of the ``direction" of the spacetime area based on the direction in which the closed spacetime contour closes (clockwise
or counterclockwise) as an extension of the right hand rule to spacetime contours and surfaces. A more rigorous way 
to determine the direction of the area for mixed spatial and time surfaces is given through the use of differential 
forms \cite{ryder, frankel, felsager}.

\end{document}